\documentstyle[12pt,aaspp4]{article}

\lefthead{Sembach, Savage, \& Hurwitz}
\righthead{Highly Ionized ISM Toward ESO\,141-055}

\begin{document}

\newcommand{\degr}{$^{\circ}$}
\newcommand{\kms}{\,km\,s$^{-1}$}     
\newcommand{\vlsr}{$v_{\sc lsr}$}

\title{GHRS and ORFEUS-II Observations of the Highly Ionized Interstellar
Medium Toward ESO\,141-055\footnotemark}

\footnotetext{Based on observations with the NASA/ESA Hubble Space Telescope, 
obtained at the Space Telescope Science Institute, which is operated by the 
Association of Universities for Research in Astronomy, Inc. under NASA 
contract NAS5-26555.}

\author{Kenneth R. Sembach}
\affil{Department of Physics \& Astronomy, The Johns Hopkins University,
3400 N. Charles St., Baltimore, MD  21218 \nl
{\it sembach@pha.jhu.edu}}

\author{Blair D. Savage}
\affil{Department of Astronomy, University of Wisconsin - Madison,
475 N. Charter St., Madison, WI  53706 \nl
{\it savage@astro.wisc.edu}}

\author{Mark Hurwitz}
\affil{Space Sciences Laboratory, University of California, Berkeley, 
CA  94720 \nl
{\it markh@ssl.berkeley.edu}}

\begin{abstract}
We present Goddard High Resolution Spectrograph and ORFEUS-II measurements
of \ion{Si}{4}, \ion{C}{4}, 
\ion{N}{5}, and \ion{O}{6} absorption in the interstellar medium
of the Galactic disk and halo toward the nucleus of the Seyfert galaxy
ESO\,141-055.  The high ionization absorption is strong, with 
line strengths consistent with the spectral
signature expected for hot (T $\gtrsim10^5$ K) collisionally ionized
gas in either a ``Galactic fountain'' or an inhomogeneous medium containing 
a mixture of conductive 
interfaces and turbulent mixing layers.  The total \ion{O}{6} column density 
of $\approx$\,10$^{15}$ cm$^{-2}$ suggests that the scale height of 
\ion{O}{6} is large ($\gtrsim$3 kpc) in this direction.
Models of the \ion{C}{4} velocity distribution 
along the sight line are consistent with a 
large scale height for the highly ionized gas.  Comparison of the high ion 
column densities
with measurements for other sight lines indicates 
that the highly ionized gas distribution is patchy.  The amount of \ion{O}{6} 
perpendicular to the Galactic plane varies by at least a factor of $\sim4$
among the complete halo sight lines thus far studied.
In addition to the high ion absorption, lines of low ionization species are 
also present in the spectra.
With the possible exception of \ion{Ar}{1}, which may have a lower than
expected abundance resulting from partial photoionization of gas 
along the sight line,
the absorption strengths are typical of those expected for the warm, neutral
interstellar medium.  The sight line intercepts a cold molecular cloud with 
$N$(H$_2$)\,$\approx$\,10$^{19}$~cm$^{-2}$.  The cloud has an 
identifiable counterpart in IRAS 100$\mu$m emission maps of this region
of the sky.  We detect a Ly$\alpha$ 
absorber associated with ESO\,141-055 at $z$\,=\,0.03492.
This study presents an enticing glimpse into the interstellar and 
intergalactic
absorption patterns that will be observed at high spectral
resolution by the Far Ultraviolet Spectroscopic Explorer.

\end{abstract}

\keywords{galaxies: individual (ESO\,141-055) -- Galaxy: halo 
-- galaxies: absorption lines -- ISM: abundances -- ISM: clouds -- 
ISM: molecules -- ultraviolet: ISM}

\section{Introduction}
	The distribution of highly ionized gas in the Galaxy contains 
information about the physical processes that transport gas and energy 
within the interstellar medium and intergalactic medium.  Absorption line 
observations of 
highly ionized atoms (e.g., \ion{Si}{4}, \ion{C}{4}, \ion{N}{5}, 
\ion{O}{6}) toward extragalactic continuum 
sources are particularly valuable since the interstellar medium (ISM) 
along entire paths through the Galaxy can be sampled and the total velocity 
extents and column densities of the absorption measured.  These quantities 
provide important benchmarks for 
comparisons with observations of the disks and 
halos of other galaxies detected through their absorption line signatures 
against distant quasars (Steidel 1990; Lu et al. 1995; Burles \& Tytler 1996;
Songaila \& Cowie 1996; Savage, Tripp, \& Lu 1998).  They also provide 
impetus for theoretical models describing the distribution and ionization 
of gas in galaxies (Spitzer 1990) and in the intergalactic medium
(Cen \& Ostriker 1999).  

Savage, Sembach, \& Lu (1997) have recently summarized most of the available 
Galactic high ion (\ion{Si}{4}, \ion{C}{4}, and N V) data obtained with
the Goddard High Resolution Spectrograph (GHRS) aboard the Hubble Space 
Telescope (HST).  The high ion species have large scale heights perpendicular 
to the Galactic plane ($h_z \sim 3-5$ kpc) and ionic ratios that are 
suggestive of multiple, collisionally ionized regions with temperatures of 
$\sim$10$^5$~K.  The highly ionized gas distribution is patchy and dependent 
upon the types of large scale interstellar structures encountered
(Sembach, Savage, \& Tripp 1997).  High quality \ion{C}{4}  and \ion{N}{5} 
absorption line data is available for approximately one dozen complete paths 
through the Galactic halo, but information for \ion{Si}{4} exists for only a 
few such sight lines (3C\,273, Mrk\,509, and a few stars in the Large and 
Small Magellanic Clouds). This situation will improve rapidly in the near 
future as extragalactic objects are observed with the Space Telescope Imaging 
Spectrograph (STIS) aboard the HST.

Information about \ion{O}{6} absorption in the Galactic halo
is also scarce.  Observations of \ion{O}{6} 
toward extragalactic sources are limited to data collected with the 
Hopkins Ultraviolet Telescope (HUT)  and the Orbiting and Retrievable Far
and Extreme Ultraviolet Spectrometers (ORFEUS) flown on two 
Space Shuttle missions.
Widmann et al. (1998) recently reported an \ion{O}{6} scale height of 
5.50$\pm^{2.37}_{2.09}$ kpc based upon moderate resolution ($\sim$0.2\AA)
ORFEUS-II echelle spectrometer
observations of one SMC star (HD\,5980) and several LMC stars (HD\,36402,
HD\,269546, LH\,10:3120, and Sk\,--67\,166).  ORFEUS-I and II  data obtained 
with the Berkeley Spectrograph for the sight lines to 3C\,273 (Hurwitz et al.
1998a) and NGC\,346 and PKS\,2155-304 (Hurwitz et al. 1995) are marginally
consistent with this result.  Eventually, the Far Ultraviolet Spectroscopic
Explorer (FUSE) will provide high-resolution \ion{O}{6} observations toward
AGNs and QSOs, but until then halo gas studies are limited to lower 
resolution data that provide reliable measures of the total amount of
 \ion{O}{6}
but do not allow for detailed velocity comparisons with the 
existing \ion{C}{4} and \ion{N}{5} data.

Obtaining complete sets of high ion column density measurements toward 
AGNs and QSOs has been a primary observational limitation to understanding the 
highly ionized ISM in the Galactic halo.  This information is necessary input 
for theories seeking to describe the production, distribution, 
and non-equilibrium cooling of hot (10$^5$--10$^6$~K) gas in the halo.  
Currently, a complete set of high ion measurements exists only for the 
3C\,273 sight line.  The present study provides information for the high 
ionization lines of \ion{Si}{4}\,$\lambda\lambda$1393.76,\,1402.77, 
\ion{C}{4}\,$\lambda\lambda$1548.19,\,1550.77, 
\ion{N}{5}\,$\lambda\lambda$1238.82,\,1242.80, and 
\ion{O}{6}\,$\lambda\lambda$1031.93,\,1037.62 toward ESO\,141-055
($l$\,=\,338.18\degr, $b$\,=\,--26.71\degr).  
ESO\,141-055 is classified in the NASA/IPAC Extragalactic Database as a 
Seyfert~1 galaxy at a redshift of 0.0371$\pm$0.0003 (de~Vaucouleurs et al. 
1991). It is bright enough at ultraviolet (UV) wavelengths to be observed with 
existing instrumentation in reasonable exposure times  
(f$_{UV} \sim 10^{-14} - 10^{-13}$ erg~cm$^{-2}$~s$^{-1}$~\AA$^{-1}$).

This study is organized as follows.  In \S2 we describe the 
observations and data reduction.  Section 3 contains the basic measurements
for the GHRS and ORFEUS-II data.  
We analyze the properties of the high ionization gas in \S4, the low 
ionization gas in \S5, and the molecular gas along the sight line in \S6.
We provide a short summary of our conclusions in \S7.

\section{Observations and Reductions}
\subsection{Goddard High Resolution Spectrograph Data}
	We used the GHRS on the HST to obtain intermediate-resolution 
spectra of the \ion{Si}{4} (1381.8--1418.6~\AA), \ion{C}{4}
(1519.8--1556.1~\AA), and \ion{N}{5} (1228.7--1265.9~\AA) spectral
regions of ESO\,141-055 in October 
1996 as part of HST Guest Observer program GO-6451.  The galaxy was 
centered in the large (1\farcs74$\times$1\farcs74) 
science aperture and observed with the G160M first-order 
grating.  Integration times were 9185 seconds for \ion{Si}{4}, 
10423 seconds for \ion{C}{4}, and 9177 seconds for 
\ion{N}{5}.  We used standard carrousel rotation (FP-Split) and 
spectrum deflection procedures to reduce fixed-pattern noise
caused by irregularities in the detector window and photocathode response.  
We used detector substepping pattern \#5 to provide full-diode array 
observations of the interorder backgrounds, which we subtracted from the 
extracted spectra.  The extracted spectra contained four substeps per 
resolution element (1 diode), which we rebinned to two samples per
 resolution element to improve the signal-to-noise ratio without 
significant loss 
of spectral resolution.  Approximate S/N levels of 13, 11, and 21 per 
resolution element were achieved for the \ion{Si}{4}, \ion{C}{4}, and 
\ion{N}{5} spectral regions, respectively.

We co-added the individual FP-Split sub-exposures using the nominal 
pixel offsets determined from the grating carrousel rotation since the signal
of the individual sub-exposures was insufficient to centroid accurately on the 
interstellar profiles.  There were 24 sub-exposures for the 
\ion{Si}{4} and \ion{N}{5} spectral regions and 28 (16+12) sub-exposures 
for the \ion{C}{4} spectral region.  These post-COSTAR\footnotemark\ 
observations have a Gaussian core containing $\approx$\,70\% of the light and 
a broad wing containing $\approx$\,30\% of the light (see Figure~4 of 
Robinson et al. 1998).
\footnotetext{COSTAR is the Corrective Optics Space Telescope Axial 
Replacement used to correct the spherical aberration in the HST primary 
mirror.}
The spectral resolution (FHWM) is approximately 14 \kms\ 
at 1550\,\AA,  16~\kms\ at 1400\,\AA, and 18 \kms\ at 1250\,\AA. 
  
The wavelength accuracy of the observations is approximately $\pm$1 diode 
(1$\sigma$), which corresponds to 18--14 \kms\ over the 1250--1550\,\AA\ 
wavelength range.  All velocities have been referenced to the Local
Standard of Rest defined by Mihalas \& Binney (1981)\footnotemark,  
where \vlsr\
= $v_{\sc helio}$ + 0.4 \kms\ for the ESO\,141-055 sight line.
\footnotetext{Their reduction to the LSR assigns a solar speed of 16.5 
\kms\ in the direction $l$\,=\,53\degr, $b$\,=\,+25\degr.}

The fully reduced GHRS data are displayed in Figure~1.  The 1$\sigma$ error
spectra are shown below each data spectrum.  Interstellar lines discussed
later are labeled.  The data are preserved in the HST archive with 
identifications of Z3170105T (\ion{Si}{4}), Z317010BT/DT (\ion{C}{4}), 
and Z3170105T (\ion{N}{5}).  Additional information about the in-orbit 
performance of the GHRS can be found in articles by
Brandt et al. (1994) and Heap et al. (1995).  Technical information about 
the instrument can be found in Soderblom et al. (1994).

\subsection{ORFEUS-II Data}
We observed ESO\,141-055 with the Berkeley spectrograph 
aboard the ORFEUS telescope during the ORFEUS-Shuttle Pallet Satellite (SPAS)
II mission in November 1996.  The exposure time for the observation was 
13828 seconds, the longest of the mission.
Background levels were monitored in an adjacent off-axis slit and were 
subtracted from the on-source data.  We produced an extracted, calibrated, 
airglow-cleaned spectrum using the standard reduction procedures and 
calibration files available at the University of California-Berkeley. 
The spectral resolution across most of the 905--1220\,\AA\ bandpass is roughly 
100 \kms\ (FWHM).  We resampled the data into 0.1\,\AA\ bins to produce a 
spectrum with S/N\,$\approx$\,7 per resolution element near 1030\,\AA. 
The absolute wavelength 
uncertainty of the data processing is roughly 0.5\,\AA, but we were able to 
improve upon this to an accuracy of $\approx$\,0.1\,\AA\ 
(30 \kms\ at 1000\,\AA) by noting the positions of 
lines in the H$_2$ Lyman series (0--0) to (8--0) vibrational bands across 
the spectrum.  

The fully reduced ORFEUS-II spectrum is shown in Figure~2.  
The far-ultraviolet continuum of ESO\,141-055 is 
relatively flat, with a rise centered near $\sim$1075\,\AA\ 
corresponding to the broad-line \ion{O}{6} emission 
from the AGN.  We have identified 
the locations of some of the interstellar atomic lines of interest in the 
figure. 
Most of the remaining unlabeled features are lines of H$_2$. 
The main diagnostic of hot gas visible in the spectrum is the \ion{O}{6} 
doublet, with lines at 
1031.93 and 1037.62\,\AA.  Inspection of the \ion{O}{1}\,$\lambda$1039.23
line near the weaker member of the \ion{O}{6} doublet confirms that the 
zero-point of the wavelength scale in the \ion{O}{6} spectral
region is consistent with that derived from the H$_2$ lines. 

For information about the design and far-ultraviolet 
performance of the Berkeley 
spectrograph, see Hurwitz \& Bowyer (1996) and Hurwitz et al. (1998b). 
Descriptions of the Astro-SPAS platform and ORFEUS project have been 
given by Grewing et al. (1991).

\section{Absorption Line Measurements}

We fit continua to the spectra shown in Figures~1 and 2 using Legendre
polynomials defined over wavelength regions free from obvious interstellar
absorption lines.  Equivalent widths and errors for the high ion lines and the 
\ion{Si}{2} and \ion{S}{2} lines present in the GHRS spectra are listed in
Table~1. We also list the central velocities and full widths at
half maximum intensity for each line.   
Discussions of continuum placement and equivalent
width determination methods can be found in Sembach \& Savage (1992).  The 
normalized line profiles are shown in Figure~3.

\subsection{The Ultraviolet Lines Between 1230\,\AA\ and 1555\,\AA}

The GHRS data are of sufficient quality to permit ion-ion profile 
comparisons as a function of velocity.   This is most readily done by 
converting the observed absorption profiles into apparent column density 
profiles in units of ions cm$^{-2}$ (\kms)$^{-1}$ through the relation 

\begin{equation}
N_a(v) = \frac{m_ec}{\pi e^2}~\frac{\tau_a(v)}{f\lambda} = 3.768\times10^{14}~\frac{\tau_a(v)}{f\lambda},
\end{equation}

\noindent
where 

\begin{equation}
\tau_a(v) = ln \frac{I_c(v)}{I_{obs}(v)}
\end{equation}

\noindent
is the apparent optical depth of the line as a function of velocity $v$.
In Eqs.(1) and (2) $I_c(v)$ and $I_{obs}(v)$ are 
the estimated continuum and observed intensities at $v$, $f$ is the 
oscillator strength of the line, and $\lambda$ is the wavelength of the line 
in\,\AA. We can check for unresolved saturated 
structure at different velocities
by comparing the profiles of multiple lines of the same species differing
in the product $f\lambda$ 
(see Savage \& Sembach 1991).  We show the $N_a(v)$ profiles for the 
\ion{S}{2}, \ion{Si}{4},
\ion{C}{4}, and \ion{N}{5} lines in Figure~4, where it can be seen that 
there is general agreement between the $N_a(v)$ profiles of the weaker line
(solid line) and the stronger line (filled circles) of each species over most
of the velocity range shown.  However, at some velocities, particularly
those near the column density peaks, the $N_a(v)$ profiles of the stronger 
lines underestimate the $N_a(v)$ profiles of the weaker lines, thereby
indicating that unresolved saturated structure exists at those velocities.    

We list the integrated apparent column densities of the ions in column~8
of Table~1.  These integrations were defined by $N_a$ = $\int$$N_a(v)$\,dv 
over the 
velocity ranges listed in column~4.  Comparison of the integrated values of 
$N_a$ for individual lines of a given species also indicates that modest
saturation corrections are needed for \ion{C}{4} and \ion{Si}{4}.  
Application of saturation corrections based upon the $N_a$
values for each species results in the final column densities listed in 
column 9.  These values were derived using the integrated values of
$N_a$ (see Savage \& Sembach 1991) and are consistent with the pixel-by-pixel
saturation correction algorithm described by Jenkins (1996).\footnotemark\

In the case of the broad, weak lines of \ion{N}{5}, we have simply averaged 
the values of $N_a$ to produce a final value of 
$N$(\ion{N}{5})\,$\approx$\,6.65$\times$10$^{13}$ cm$^{-2}$.  It is 
unlikely that there are narrow components with $\tau$~$\gg$~1 in these
profiles.  The ionization potential for the creation of \ion{N}{5} is 
77.5 eV, and most of the \ion{N}{5} is probably collisionally ionized 
in hot (T $\gtrsim10^5$ K) gas.

\footnotetext{The S/N  of our data is too low to trust 
saturation corrections derived 
from the Jenkins (1996) method for individual pixels, but
an integration of the corrected profiles yields a result similar
to the one obtained by applying the Savage \& Sembach (1991) correction.  
This is expected since the 
saturation corrections are modest.}

We note the presence of weak \ion{Mg}{2}\,$\lambda\lambda$1239.925, 1240.395
absorption near \ion{N}{5}.  These lines are indicated in Figures~1 and 3.
There may also be a small amount (W$_\lambda < 150$~m\AA) 
of CO (A--X)$_{0-0}$ band absorption
near 1544.5\,\AA.  Combined with W$_\lambda < 70$~m\AA\ for the 
(A-X)$_{5-0}$ band near 1392.5\,\AA, we set a 2$\sigma$ limit of 
log $N$(CO) $\lesssim$ $4\times10^{14}$ cm$^{-2}$.
The only other noticeable absorption feature 
in the GHRS spectra is a redshifted 
Ly$\alpha$ line at 1258.126\,\AA\ ($z$\,=\,0.03492).  This line has 
W$_\lambda$(Ly$\alpha$)\,$\approx$\,220\,m\AA, which places it
on the flat part of the curve-of growth.  The absorber is probably associated 
with ESO\,141-055 given the similarity in the redshifts of the two objects.
 
\subsection{The Far Ultraviolet Lines}

The far-ultraviolet spectrum of ESO\,141-055 shown in Figure~2 reveals a 
rich set of absorption lines due to both atomic and molecular species.  
At the spectral resolution of the ORFEUS-II data ($\approx$\,100 \kms), 
many of the features, including the \ion{O}{6} lines, are blended with
nearby lines.  To determine the possible effects of this blending on the 
amount of \ion{O}{6} derived for the sight line, we have 
replotted the 1000--1100\,\AA\ ORFEUS-II data in Figure~5.  Atomic lines near 
the \ion{O}{6} lines include \ion{C}{2}\,$\lambda$1036.34, 
\ion{C}{2}$^*$\,$\lambda$1037.02, and multiple lines in the
(6--0) and (5--0) vibrational bands of H$_2$ between 1031.5--1032.5\,\AA\ and 
1036.5--1038.7\,\AA\ (Morton 1991; Sembach 1999). 

We created a simple model of the neutral gas
absorption along the sight line.  
Using the GHRS \ion{Si}{2} and \ion{S}{2}
line profiles as a template for the neutral atomic 
gas velocity distribution, we simulated 
the absorption with the 
following parameters: $N$(\ion{H}{1}) = 3.5$\times$10$^{20}$ cm$^{-2}$, b =
28 \kms, and \vlsr\ = 0 \kms.  The model does not include ionized gas.
The total \ion{H}{1} column density of 3.5$\times$10$^{20}$ cm$^{-2}$
provides reasonable agreement with the observed Ly$\beta$ absorption,
the amount of \ion{H}{1}
expected for a low latitude extragalactic sight line (Savage et al. 1997),
the approximate S/H ratio for a halo sight line
[$N$(\ion{S}{2})/$N$(\ion{H}{1})\,$\approx$\,(0.5--1.0)\,$\times$($N$(S)/$N$(H))$_\odot$] (Savage \& Sembach 1996)\footnotemark,
\footnotetext{The column density of \ion{S}{1} along the sight line is small,
log $N$(\ion{S}{1}) $<$ 14.09 (Table~1).  Thus, $N$(\ion{S}{1})/$N$(\ion{S}{2})
$<$ 0.04.}
and the \ion{H}{1} column of $\sim$5$\times$10$^{20}$ cm$^{-2}$ derived 
from the \ion{H}{1} 21\,cm emission data of Colomb, Poppel, \& Heiles
(1980)\footnotemark.
\footnotetext{The half-power beam width for the 21\,cm radio
observations is 30\arcmin.
Therefore, there may be additional \ion{H}{1} emission contributing to
the 21\,cm profile from
regions near to, but not directly along, the sight line sampled by
the absorption measurements (see \S6).}  
The model is represented by the heavy solid line overplotted on the 
spectrum in Figure~5. At the resolution of the data, 
the precise details of the model velocity distribution do not strongly
affect our primary conclusions.  
The absorption lines in the model were constructed using solar reference 
abundances (Anders \& Grevesse 1989; Grevesse \& Noels 1993) and atomic data 
compiled by Morton (1991).

We have also included a molecular component in the model shown in Figure~5
to reproduce the numerous H$_2$ lines observed.  A single component with 
$N$(H$_2$) = 1$\times$10$^{19}$ cm$^{-2}$, b = 10 \kms, \vlsr\ = 0 \kms, and 
T$_{rot}$ = 100 K reproduces
many of the observed absorption features well.  The column density and 
rotational temperature of the molecular component can be varied by roughly 
$\pm^{2.0}_{0.5}\times$10$^{19}$ cm$^{-2}$ and $\pm$20\degr\ K, respectively, 
before the fits become noticeably discrepant from the data.  Lower temperature 
models underproduce the amount of H$_2$
in the J\,$\ge$\,2 levels, while higher temperature models
overproduce higher J level absorption.   In the adopted model, 
38\% of the H$_2$ is in the J = 0 level, 61\% is in the J = 1
level, and 1\% is in the J = 2 level.  Adopting a lower b-value for the 
molecular gas requires
a higher value of $N$(H$_2$) to match the data. Higher resolution data are 
needed to determine the velocity distribution of the molecular gas and 
reliable column densities in each rotational level.

It is clear from the model shown in Figure~5 that atomic lines of 
\ion{Si}{2}, \ion{Ar}{1}, and \ion{Fe}{2} should be strong unless the elements 
are depleted from the gas onto dust or are ionized into higher ionization 
stages. Of special note is the apparent absence of strong 
absorption by \ion{Fe}{2} and \ion{Ar}{1}.  Lines of these species
are indicated with asterisks below the spectrum in Figure~5.  We discuss 
these lines further in \S5.

Of the two members of the \ion{O}{6} doublet, the 1031.93\,\AA\ line
is less susceptible to blending with other lines and is strong in this 
spectrum.
Note the possible presence of the 1031.19\,\AA\
(6--0) P(3) line of H$_2$ in the blue wing of the \ion{O}{6}
line at a velocity of 
--214 \kms.  Integrating over the --150 to +100 \kms\ velocity range, we find
W$_\lambda$(1031.93)\,$\approx$\,537$\pm$100 m\AA, 
which converts to a 
logarithmic column density log\,$N$(\ion{O}{6})\,=\,15.1$\pm^{0.6}_{0.3}$, 
assuming the absorption
can be described by a single component with a width b\,$\approx$\,57$\pm$8 
\kms\ derived by fitting a single Gaussian component to the \ion{N}{5} lines. 
A velocity component distribution having several narrower components resembling
those seen in the \ion{C}{4} profiles produces a similar result.
\ion{O}{6}\,$\lambda$1037.62 is severely blended with strong 
\ion{C}{2} absorption centered near --370 \kms\ and the R(0), R(1), 
and P(1) lines of the (5--0) Lyman series vibrational band of H$_2$ 
centered at --311, --135, and +157 \kms, respectively.
We do not consider the
\ion{O}{6} 1037.62\,\AA\ line further in this study.  

We find only marginal evidence
for Ly$\beta$ absorption at 1061.54\,\AA\ 
associated with the $z = 0.03492$ Ly$\alpha$
absorber seen in our GHRS spectrum.  For this absorber, we estimate that 
W$_\lambda$(Ly$\beta$)\,$\approx$\,120$\pm$70 m\AA, which together with the 
Ly$\alpha$ measurement (\S3.1) suggests a total column density of 
log\,$N$(\ion{H}{1}) = 13--16.
The unidentified absorption feature near 1041.2\,\AA\ is unlikely to 
be a Ly$\beta$ line from an intergalactic cloud at redshift 
$z = 0.0151$ since there
is no corresponding Ly$\alpha$ absorption at 1234.0\,\AA\ in Figure~1
(W$_\lambda$ $<$ 40 m\AA, $N$(\ion{H}{1}) $<$ 7.3$\times$10$^{12}$ cm$^{-2}$).

\section{High Ionization Gas}

\subsection{Scale Heights}
The integrated column densities of the high ionization species observed
toward ESO\,141-055 are consistent with an extended distribution of 
highly ionized gas away from the Galactic plane.  Assuming an exponential
density distribution given by 
n(z) = n$_0$\,e$^{-|z|/h_z}$, the integrated
column density of an ion in a direction perpendicular to the Galactic disk is
$N_\perp$(ion) = $N$(ion)\,sin$|b|$ = n$_0h_z$.  Values of $N_\perp$ for the
highly ionized species toward ESO\,141-055 are listed in Table~2 along with
values for the sight lines to HD~36402 (in the LMC), PKS~2155-304, and 
3C~273.  Table~2 also contains the results of fitting exponential distributions
to the values of $N_\perp$ for \ion{Si}{4}, \ion{C}{4}, and \ion{N}{5} 
for an ensemble of sight lines toward $\sim$10 AGN/QSOs and $\sim$30 OB
stars (Savage et al. 1997).  The ESO\,141-055 values are systematically 
higher than the Galactic averages by $\sim$0.1--0.3 dex, but are less than
those toward 3C~273, which may have a substantial high ion
contribution arising within Galactic Radio Loops I and IV 
(see Sembach et al. 1997).  

Values of $N_\perp$(\ion{O}{6}) have been measured for only a few sight lines
through the Galactic halo.  The directions studied include 4 LMC and 2 SMC 
sight lines
(Hurwitz et al. 1995; Widmann et al. 1998), PKS~2155-304 (Hurwitz et al.
1995), and 3C~273 (Davidsen 1993; Hurwitz et al. 1998a).  This has made it 
difficult to determine an \ion{O}{6} scale height that can be generalized
to the Galaxy as a whole.  Indeed, values
of $h_z$ ranging from a low of $\sim$100 parsecs (Jenkins 1978a,b; Hurwitz 
\& Bowyer 1996) up to $\sim$5 kiloparsecs (Widmann et al. 1998) have been 
reported in the literature.  
As more data become available both for extended sight lines in the Galactic
disk and sight lines through the Galactic halo, it will be possible to place
better constraints on n$_0$ and $h_z$.  The results for 
ESO\,141-055 favor a scale height on the order of the scale heights found by
Savage et al. (1997) for the other high ions: 
$h_z$(\ion{Si}{4})\,$\approx$\,5.1 kpc,
 $h_z$(\ion{C}{4})\,$\approx$\,4.4 kpc, 
and  $h_z$(\ion{N}{5})\,$\approx$\,3.9 kpc.  Assuming
n$_0$(\ion{O}{6}) $\approx$\,1--2$\times$10$^{-8}$ ions cm$^{-2}$ from
local estimates of $N$(\ion{O}{6}) made with the Copernicus satellite (Shelton
\& Cox 1994), the ESO\,141-055 value of 
$N_\perp$(\ion{O}{6})\,$\approx$\,4.5$\times$10$^{14}$ cm$^{-2}$
implies $h_z$(\ion{O}{6}) $\approx$\,7--14 kpc.  If n$_0$(\ion{O}{6}) is as 
high as 5$\times$10$^{-8}$ ions cm$^{-2}$ (Hurwitz \& Bowyer 1996), then 
the inferred value of  $h_z$(\ion{O}{6}) becomes $\approx$3 kpc.
	
Patchiness in the gas distribution complicates scale height 
determinations and necessitates determinations of $N_\perp$ for an ensemble of
sight lines before meaningful estimates of $h_z$ and its error
can be made and applied to more general problems.
Several lines of evidence point to a clumpy distribution for the 
\ion{O}{6}-bearing gas in the Galactic disk and halo.  The values of 
$N_\perp$(\ion{O}{6}) in Table~2 indicate that there is at least
a  factor of $\sim$4 range in the \ion{O}{6} columns perpendicular
to the disk in the inner regions
of the Galaxy.  This spread is comparable to the patchiness found for 
\ion{C}{4} and \ion{N}{5} at many distances from the Galactic plane over 
large angular scales (Savage et al. 1997).  Over small angular separations, the
limited data suggest that $N_\perp$(\ion{O}{6}) may vary more than 
$N_\perp$(\ion{C}{4}). For example, $\Delta$$N_\perp$(\ion{O}{6}) is 
at least a factor of 3.7 between LMC and SMC directions, compared to 
$\Delta$$N_\perp$(\ion{C}{4}) $\approx$\,2.0.  Additional evidence for 
patchiness of the \ion{O}{6} distribution comes from a study of 
ORFEUS-I spectra for halo stars  by Hurwitz \& Bowyer (1996),
who found low columns of \ion{O}{6} out to $|$z$|$ $<$ 3 kpc.

ESO\,141-055 intercepts Radio Loop~I near its southern boundary
(Berkhuijsen, Haslam, \& Salter 1971).  
The large amount of high ion gas observed may be due in part to 
passage of the sight line through this structure (see Sembach et al. 1997), 
but some of the \ion{O}{6}
absorption could be associated with more distant regions.  
The
sight line also passes through the 3/4 keV X-ray enhancement attributed to the 
Galactic bulge (Snowden et al. 1995).
Observations of \ion{O}{6} along other sight lines in this region of the
sky will be needed to separate foreground absorption produced by Loop~I
from the contributions of more distant regions.

\subsection{Velocity Structure}

The high ionization lines of \ion{Si}{4} and \ion{C}{4} (and to a 
lesser degree \ion{N}{5}) have
a velocity structure indicative of multiple components
over the LSR velocity range from --130 to +140 \kms.  Distinct features
can be seen at 
\vlsr\ $\approx$ --60, +15, and +45 \kms\ (Figure~3). 
The profiles also show a positive velocity wing extending to $\approx$\,+120
\kms.  In this direction, differential Galactic rotation can result in 
negative line-of-sight velocities out to a distance of 17 kpc (z = --8 kpc). 
We show the rotation curve for a co-rotating gas distribution 
along the sight line in the top right panel of Figure~6.  
The minimum predicted velocity of --103 \kms\ occurs at d = 8.8 kpc 
(z = --4.0 kpc, R$_g$ = 3.1 kpc).\footnotemark 

\footnotetext{R$_g$ is the distance from the Galactic center of the projection
of the sight line onto the Galactic plane.}

The ESO\,141-055 sight line passes through the inner regions of the Galaxy
at a minimum galactocentric radius of 3.1 kpc, where non-circular motions may
confuse the absorption signature expected for gas participating in 
differential 
Galactic rotation.  Previous studies of the high ion absorption toward inner 
Galaxy stars in this general direction (Savage, Massa, \& Sembach 1990;
Sembach, Savage, \& Massa 1991; Sembach, Savage, \& Lu 1995; 
Savage, Sembach, \& Cardelli 1994)
have shown that the highly ionized ISM is complex, with kinematics 
varying as a function of distance from the Galactic plane.  Possible 
explanations for the presence of ``forbidden velocity'' gas in the inner
Galaxy have been summarized by Tripp, Sembach, \& Savage (1993). 

To determine the nature of the gas distribution along the ESO\,141-055
sight line, we have constructed several simple models for a smoothly 
distributed gas having an exponentially declining density with distance
from the Galactic plane.  We have created the apparent column density profiles
expected for several values of the intrinsic gas velocity
dispersion ($\sigma$~=~10, 20, 30 \kms) and gas scale height ($h_z$~=~1, 5, 10
kpc).  These profiles are shown in Figure~6 along with the \ion{C}{4}
$N_a(v)$ profiles for the ESO\,141-055 sight line and for four sight lines
toward stars in the inner Galaxy.  The locations of the sight lines are 
shown schematically in Figure~7.  Several conclusions can be drawn from
a direct comparison of the ESO\,141-055 profile to the other profiles.

\noindent
(1) None of the stellar sight lines shows the strong peak in \ion{C}{4} 
near +15 \kms\ seen toward ESO\,141-055.  This feature cannot be reproduced
in the smooth gas model profiles without causing serious disagreement at 
other velocities.
The high ion gas at these velocities might be associated with the low ion
(and perhaps molecular) gas near +5 \kms. As expected, the 
ESO\,141-055 sight line exhibits less total \ion{C}{4} absorption than the 
extended, lower latitude sight lines shown in Figure~6.

\noindent
(2) The ESO\,141-055 sight line has a \ion{C}{4} extension
to negative velocities comparable to 
that of the HD\,151990, HD\,156359, and HD\,163522 sight lines.  However,
the positive velocity wing is seen only toward the two stars located at 
smaller Galactic longitudes (HD\,151990 and HD\,156359).  
The smooth gas distribution models shown in 
the right panel of Figure~6 indicate that substantial column densities
at positive 
velocities are expected for corotating gas only if the scale height is
large ($h_z$ $>$ 5 kpc).  The presence of the positive velocity 
wing toward HD\,151990, which attains a maximum z-distance of 430 pc, 
suggests that the positive velocity 
absorption toward ESO\,141-055 may actually occur
within a few hundred parsecs of the disk.  Alternatively, 
if deviations from corotation are present in the low halo
as suggested by Sembach et al. (1991) for the HD\,156359 sight line, the
absorption may occur at higher altitudes (z = 1--2 kpc).

\noindent
(3) The smooth gas distribution models shown in Figure~6 can roughly 
reproduce the observed \ion{C}{4} $N_a(v)$ profile toward ESO\,141-055
if the strong absorption near +15 \kms\ is attributed to an additional
enhancement along the sight line.  The best fit parameters ignoring the 
+15 \kms\ feature
are n$_0$(\ion{C}{4}) = 4.0$\times$10$^{-9}$
cm$^{-2}$, $h_z$(\ion{C}{4}) = 11.5 kpc, and $\sigma$(\ion{C}{4}) = 30 \kms.
Similar values of $h_z$ and $\sigma$ are obtained when the \ion{N}{5} $N_a(v)$
profiles are modeled.
The major shortcoming of this model is the overprediction in the amount
of \ion{C}{4} and \ion{N}{5} at \vlsr\ $<$ --80 \kms.
While the high ion scale height derived from this kinematical analysis
is roughly consistent with the large value inferred from the $N\sin |b|$
considerations for \ion{O}{6} (\S4.1), such a large scale
height is out of character when other sight lines through the Galactic 
center are considered (see Savage et al. 1997).   We conclude that there is
either an enhancement in the high ion columns along the ESO\,141-055 sight 
line compared to these other inner Galaxy sight lines, or that the global 
distribution of gas in the region of the Galactic center is irregular.  In
either case, the highly ionized gas distribution is patchy in nature, so it 
is unwise to draw Galaxy-wide conclusions from the analysis of 
observations of a small number of lines of sight.

\subsection{Ionization}

Of the ions observed in this study, \ion{O}{6} is the best tracer of
gas at coronal temperatures.  It peaks in abundance in 
collisional ionization equilibrium at $\sim$2.8$\times$10$^5$ K, compared
to 1.8$\times$10$^5$ K, 1.0$\times$10$^5$ K, and 6.3$\times$10$^4$ K
for \ion{N}{5}, \ion{C}{4}, and \ion{Si}{4}, respectively 
(Sutherland \& Dopita 1993).  Furthermore, the predicted strengths of 
the \ion{O}{6} lines are often several times those of the 
\ion{C}{4} and \ion{N}{5} lines.  \ion{O}{6} traces gas at temperatures up to 
$\sim$10$^6$~K and is not produced in significant quantities by 
photoionization in the ambient ISM (Benjamin \& Shapiro 1999).

The velocity resolution of our ORFEUS-II data 
does not allow us to determine the \ion{O}{6} column densities of individual 
absorbing regions along the ESO\,141-055 sight line.
However, it is possible to compare the integrated strength of the \ion{O}{6}
absorption to the other high ions and
derive information about the sight-line-averaged ionization conditions in
this direction.  The high ion ratios 
are listed in Table~3, where they are compared to 
values for other sight lines and 
to theoretical predictions. The high ion ratios toward ESO\,141-055 are 
marginally higher than those observed toward 3C\,273.  

The $N$(\ion{C}{4})/$N$(\ion{O}{6}) ratio is suggestive of a mix of disk and
halo gases along the ESO\,141-055 sight line.  The observed value of 
0.55$\pm$0.30 
is near the high end of values found for the Galactic disk and the low end of 
values for the Galactic halo (Spitzer 1996). A 
similar conclusion is reached if \ion{N}{5} and \ion{O}{6} are
compared.  We find $N$(\ion{O}{6})/$N$(\ion{N}{5})\,$\approx$\,15, which is
the same as the average value found for the Galactic disk by  
Hurwitz \& Bowyer (1996).

The radiatively cooling fountain flow model proposed by Benjamin \& Shapiro
(1999) to explain the presence of highly ionized gas in the Galactic halo
provides a reasonable description of the integrated high ion ratios toward
ESO\,141-055 (see Table~3).  A 
combination of turbulent mixing layers (Slavin, Shull, \& Begeleman 1993)
and conductive interfaces (Borkowski, Balbus, \& Fristrom 1990) in the 
right proportions could also reproduce the observed values.  
Evidence for an interaction of hot and cool (or warm) gases is provided by the 
apparent column density profiles shown in Figure~4.  The high ion profiles 
have a noticeable peak near \vlsr\,$\approx$\,+15 \kms, which is near the
peak of the \ion{S}{2} profiles.  The observed ratio 
$N_a$(\ion{C}{4})/$N_a$(\ion{Si}{4}) $\approx$ 4--5 is roughly constant over 
the --100 to +100 \kms\ velocity range, but is likely to 
be somewhat higher near --60 \kms\ and +15 \kms\ due to modest
saturation corrections required for \ion{C}{4} at those velocities. Hybrid
descriptions of the hot gas distribution in the Milky Way have been 
suggested to explain the high ion observations (e.g., Shull \& Slavin 1994; 
Spitzer 1996), but it is difficult to distinguish among the many possible 
combinations without a significantly larger data set for \ion{O}{6} than
now exists.  We anticipate that FUSE will fill this vital
need and shed new light on the issue of the ionization of the disk and
halo gases.

\section{Low Ionization Gas}

The amount of high-resolution information for the low ionization 
interstellar gas toward ESO\,141-55 is limited.  
The \ion{Mg}{2}\,$\lambda$1240 doublet lines are weak -- W$_\lambda \approx 50$ m\AA, log $N$(\ion{Mg}{2}) $\sim 5\times10^{15}$ cm$^{-2}$.
The absorption lines of \ion{Si}{2} and \ion{S}{2} attain maximum depth 
near $\sim$+5 \kms.  The 
\ion{S}{2} lines have a weak, negative velocity wing that could be associated
with some of the higher ionization gas (see Figure~3). 
The overall velocity extents of the \ion{Si}{2}\,$\lambda$1260.42 line
and the high ion lines are similar.   However, unlike
the high ion profiles, which are stronger at --50 \kms\ than at +50
\kms, \ion{Si}{2}\,$\lambda\lambda$1260.42, 1526.71 exhibit much 
stronger absorption at positive velocities than the high ions. Note 
that \ion{Si}{2}\,$\lambda$1260.42 includes blended 
\ion{Fe}{2}\,$\lambda$1260.53 absorption centered near +26 \kms. 

Several conclusions can also be drawn from a comparison of the ORFEUS-II data 
and the model fit shown in Figure~5.  
First, atomic lines of \ion{H}{1} (Ly$\beta$), 
\ion{C}{2}, \ion{O}{1}, \ion{Si}{2}, \ion{Ar}{1}, and \ion{Fe}{2} should be 
present unless the elements are depleted from the gas onto dust or 
ionized into higher ionization stages.  \ion{H}{1}, \ion{C}{2}, \ion{O}{1}, 
and \ion{Si}{2} are all clearly detected.  We find 
W$_\lambda$(\ion{Si}{2}\,$\lambda$1020.7) $\approx$ 380$\pm$140 m\AA\ and 
W$_\lambda$(\ion{O}{1}\,$\lambda$1039.2) $\approx$ 470$\pm$105 m\AA. 
Absorption by \ion{Fe}{2} or 
\ion{Ar}{1} is not as strong as predicted by the model.  Fe is
easily incorporated into grains, and even in the warm neutral ISM of the 
Galactic halo is depleted by a factor of $\sim$5 (see Savage \& Sembach 
1996).  Depletion factors of refractory elements in the warm ionized ISM
appear to be comparable (Howk \& Savage 1999).  Thus, the absence of gas-phase
\ion{Fe}{2} could be explained by the presence of dust.
However, the absence of \ion{Ar}{1} absorption cannot be so readily explained
by depletion effects since  Ar is a noble element, and its abundance is 
probably not affected significantly by depletion (see Jenkins 1987). 
Sofia \& Jenkins (1998) considered the ionization
of Ar by starlight and found that \ion{Ar}{1}
is highly susceptible to photoionization in low-density, partially 
ionized regions.  
For a dilute gas that is approximately half ionized, they 
estimated that $N$(\ion{Ar}{1})/$N$(\ion{H}{1}) 
can be as small as 0.1 times the solar value of $N$(Ar)/$N$(H).  

If we assume that all of the \ion{Ar}{1} and \ion{S}{2} along the sight line  
arises in \ion{H}{1} regions and 
$N$(\ion{Ar}{1})/$N$(\ion{S}{2})\,$\approx$\,($N$(Ar)/$N$(S))$_\odot$ = 0.195, 
then we 
expect log\,$N$(\ion{Ar}{1})\,$\approx$\,14.74.  For single absorption
components with 
b-values of 5, 10, and 30 \kms, this column density corresponds to equivalent
widths of 72, 131, and 323 m\AA\ for the \ion{Ar}{1} 1048.2\,\AA\ line.
(The model line shown in Figure~5 has W$_\lambda$ = 363 m\AA.) Integrating
over the $\pm$200 \kms\ velocity range, we find 
W$_\lambda$(\ion{Ar}{2}\,$\lambda$1048.2) $<$ 200 m\AA.  There could be a 
substantial amount of \ion{Ar}{1} hiding
in the ORFEUS-II spectrum provided that the b-value of the \ion{Ar}{1}
is lower than about 20 \kms.  Partial photoionization of the gas would
reduce the expected amount of \ion{Ar}{1} relative to \ion{S}{2} since
\ion{Ar}{1} traces mainly \ion{H}{1} gas, whereas \ion{S}{2} traces both 
\ion{H}{1} and \ion{H}{2} region gas.

Trying to resolve this ionization issue would be valuable since the 
nature of ionizing sources in the Galactic halo is still poorly understood. 
The sight line is confined to the Galactic disk ($|$z$|$ $<$ 500 pc) for only 
$\sim$1 kpc.    H$\alpha$ measurements
reveal that the diffuse ionized gas in the Galaxy, which has a scale height of 
$\sim$1 kpc and a filling factor of roughly 20\%, is probably photoionized
by photons from O stars (Reynolds 1993).  Ions in the 900--1200\,\AA\
wavelength range that could be used to study the ionization conditions
include N\,{\sc i-ii}, P\,{\sc ii-v}, Ar\,{\sc i-ii}, and 
Fe\,{\sc ii-iii}.   The S/N and spectral resolution
of our ORFEUS-II data do not allow these types of comparisons.

\section{Molecular Gas}

It is generally believed that extended sight lines through the Galactic halo 
are
low density paths containing little cold atomic or molecular gas.  It is
therefore interesting that the far-ultraviolet spectrum of ESO\,141-055
shown in Figures~2 and 
5 shows clear evidence of absorption due to H$_2$.  The average
\ion{H}{1} density is roughly 0.11 atoms cm$^{-2}$, assuming  
$N$(\ion{H}{1}) = 3.5$\times$10$^{20}$ atoms cm$^{-2}$ and a 1 kpc path
length.  The H$_2$ column
density is roughly 1$\times$10$^{19}$ molecules cm$^{-2}$, which corresponds
to a sight-line-averaged molecular gas fraction 
f$_{H_2}$\,=\,2$N$(H$_2$)\,/\,[2$N$(H$_2$)+$N$(\ion{H}{1})]
$\approx$ 0.05.  This value of f$_{H_2}$ is typical of the sight lines with 
color excesses E(B-V)\,$\approx$\,0.1 mag surveyed by Savage et al. (1977).
Most of the molecular gas likely arises within a single component
since a column density of $N$(H) $\sim$ 5$\times$10$^{20}$ 
cm$^{-2}$ is needed to effectively self-shield the H$_2$ from the 
interstellar UV radiation field (Savage et al. 1977).   The H$_2$ absorption
probably occurs at a velocity near the peak of the \ion{S}{2} absorption 
shown in Figure~4.  The marginal detection of CO near 1544.5\,\AA\ (Figure~1,
\S3.1) is qualitatively consistent with such a velocity.

A possible site for the H$_2$ absorption is shown in Figure~8.  
The sight line lies within an enhancement in the dust distribution
revealed by 100$\mu$m emission in the 
IRAS Sky Survey Atlas (Wheelock et al. 1994).  The 
100$\mu$m feature is approximately 1\degr\ across and appears loosely
connected to similar features in this region of the sky.  The intensity of the 
100$\mu$m emission along the sight line is 5.0 MJy sr$^{-1}$, which
exceeds the intensity of 88\% of the pixels in the 7.5\degr$\times$7.5\degr\ 
field shown.  Note that the intensity of the emission changes rapidly over
angular scales of 10--15 arc minutes. Using the approximate relation 
$<$$I_{100}/N$(H)$>~\approx (1.0\pm0.3)\times10^{-20}$ MJy~sr$^{-1}$ cm$^2$ 
appropriate for the infrared cirrus (Reach, Koo, \& Heiles 1994),
we find that the 100$\mu$m emission in this feature corresponds to 
$N$(H)~$\approx (3.5-6.5)\times10^{20}$ cm$^{-2}$. This feature 
is probably an important source of \ion{H}{1} as well as H$_2$ along the 
sight line.  The distance of the cloud is unknown, but typically
molecular clouds are within $\sim$300 pc of the disk (Dame \& Thaddeus 
1994), which translates to a distance of $<$700 pc in this direction. 

The H$_2$ gas along the ESO\,141-055 sight line can be modeled as a 
relatively cold (T$_{rot}$\,$\approx$\,100 K) cloud (see \S3.2).  
Non-equilibrium population of the rotational levels may also be
important, especially if strong sources of UV radiation exist
in the halo. ORFEUS-I
observations of H$_2$ toward several early type stars located at distances of 
1--4 kpc within the Galactic disk revealed H$_2$ column densities about a 
factor of 10 higher than the H$_2$ column density along the ESO\,141-055 
sight line.  Those clouds have similar temperatures 
(T$_{rot}$\,$\approx$\,80--100 K) and molecular hydrogen fractions 
(f$_{H_2}$\,$\approx$\,0.06--0.16) (Dixon, Hurwitz, \& Bowyer 1998).  Thus,
the overall amount and general properties of the H$_2$ toward ESO\,141-055 are
consistent with what is expected for an extended, moderate latitude sight line.
In this context, we note that ORFEUS-II observations of the 3C~273 sight
line, which is located at $b$ = +64.4\degr, revealed at modest statistical
significance a much lower column of H$_2$ ($\sim$10$^{15}$ cm$^{-2}$)
 (Hurwitz et al. 1998a).  Further determinations of the properties
of the molecular gas toward ESO\,141-055 and along other halo sight lines
would provide an important benchmark for studies of H$_2$ in the Magellanic
Clouds (Richter et al. 1998, de~Boer et al. 1998) and quasar absorption
line systems (Foltz, Chaffee, \& Black 1988; Ge \& Bechtold 1997).

\section{Conclusions }
We present GHRS and ORFEUS-II absorption line data of the high 
ionization lines of \ion{Si}{4}, \ion{C}{4}, \ion{N}{5}, and \ion{O}{6}
in the ISM of the Galactic disk and halo toward ESO\,141-055 
($l$\,=\,338.18\degr, $b$\,=\,--26.71\degr).  We find that the high ionization 
absorption is strong, with integrated high ion column density ratios 
consistent with the signature expected for radiatively cooling gas in a
``Galactic fountain'' or an inhomogeneous ISM containing conductive 
interfaces and turbulent mixing layers.   The \ion{O}{6} column density 
perpendicular to the 
Galactic plane, $N$sin$|b|$ $\approx$\,4.5$\times$10$^{14}$ cm$^{-2}$,
suggests that the scale height of \ion{O}{6} is large 
($\gtrsim$3~kpc),  but further work is needed to generalize this result to
other regions of the Galaxy.   Models of the velocity distribution of the 
\ion{C}{4} absorption toward ESO\,141-055 are also consistent with a large 
high ion scale height, though they cannot reproduce all of the structure 
observed in the \ion{C}{4} profiles.  Some of the high ion absorption along 
the sight line may occur in the Galactic bulge region, which emits strongly 
at soft X-ray wavelengths.
Comparison of the ESO\,141-055 high ion column densities with those of
other sight lines indicates that the \ion{O}{6} distribution is at 
least as patchy as that of \ion{C}{4}.  

Low ionization absorption lines are also detected toward ESO\,141-055.
The strengths of these lines in the ORFEUS-II spectrum are difficult to gauge
with the available data, but it appears that \ion{Ar}{1} is considerably
weaker than expected.  The 
sight line intercepts a cold molecular cloud containing $\approx$\,10$^{19}$ 
molecules cm$^{-2}$ of H$_2$, which produces easily identifiable (and 
somewhat confounding)
blends of absorption at the 100\kms\ resolution of the ORFEUS-II data.  
High-resolution FUSE observations of this sight line 
would be of considerable value in sorting out the velocity 
structure and physical conditions in the molecular, cold atomic, and highly
ionized ISM along the sight line.

\acknowledgments  
We thank Steve Federman for useful comments on the manuscript.  We 
acknowledge use of the SkyView data browser available on the internet
at the Goddard Space Flight Center (http://skyview.gsfc.nasa.gov/skyview.html).
KRS acknowledges support from NASA Long Term Space Astrophysics grant 
NAG5-3485 and from the FUSE project through grant NAS5-32985.  
BDS appreciates support from NASA through grant GO-06451.01-95A from the 
Space Telescope Science Institute.  MH acknowledges support from NASA grant 
NAG5-696.
 
\references

\reference{}Anders, E. \& Grevesse, N.  1989, Geochim. Cosmochim. Acta, 53, 197
\reference{}Benjamin, R. \& Shapiro, P.  1999, ApJS, submitted
\reference{}Berkhuijsen, E., Haslam, C.G.T., \& Salter, C.J.  1971, 
A\&A, 14, 252
\reference{}Borkowski, K.J., Balbus, S.A., \& Fristrom, C.C.  1990, ApJ, 
355, 501
\reference{}Brandt, J.C., et al.  1994, PASP, 106, 890
\reference{}Burles, S. \& Tytler, D.  1996, ApJ, 460, 584
\reference{}Cen, R. \& Ostriker, J.P. 1999, ApJ, 514, 1
\reference{}Cleary, M.N., Heiles, C., \& Haslam, C.G.T.  1979, A\&AS, 36, 95
\reference{}Clemens, D.P.  1985, ApJ, 295, 422
\reference{}Colomb, F.R., Poppel, W.G.L., \& Heiles, C.  1980, A\&AS, 40, 47
\reference{}Dame, T.M. \& Thaddeus, P.  1994, ApJ, 436, L173
\reference{}Davidsen, A.F.  1993, Science, 259, 327
\reference{}de~Boer, K.S., Richter, P., Bomans, D.J., Heithausen, A., \& 
Koorneef, J.  1998, A\&A, L5
\reference{}de~Vaucouleurs, G., de~Vaucouleurs, A., Corwin, H.G. Jr., Buta, 
R.J., Pautrel, G., \& Fougue, P.  1991, Third Reference Catalogue of Bright 
Galaxies, v3.9
\reference{}Dixon, W.V., Hurwitz, M., \& Bowyer, S.  1998, ApJ, 492, 569
\reference{}Dufton, P.L., Hibbert, A., Kingston, A.E., \& Tully, J.A. 1983,
MNRAS, 202, 145-50
\reference{}Federman, S.F. \& Cardelli, J.A.  1995, ApJ, 452, 269
\reference{}Foltz, C.B., Chaffee, F.H., \& Black, J.H.  1988, ApJ, 324, 267
\reference{}Ge, J \& Bechtold, J.  1997, ApJ, 477, L73
\reference{}Grevesse, N. \& Noels, A.  1993, in Origin of the Elements, eds.
N. Prantzos, E. Vangioni-Flam, \& M. Cass\'e (Cambridge: Cambridge University 
Press), 15
\reference{}Grewing, M., et al.  1991, in Extreme Ultraviolet Astronomy, eds. 
R.F. Malina \& S. Bowyer (Elmsford: Pergamon), 437
\reference{}Heap, S.R., et al.  1995, PASP, 107, 871
\reference{}Howk, J.C. \& Savage, B.D.  1999, ApJ, in press
\reference{}Hurwitz, M. \& Bowyer, S.  1996, ApJ, 465, 296
\reference{}Hurwitz, M., Bowyer, S., Kudritski, R.P., \& Lennon, D.J. 1995, 
ApJ, 450, 149
\reference{}Hurwitz, M., Appenzeller, I., Barnstedt, J., Bowyer, S., Dixon, 
W.V., et al.  1998a, ApJ, 500, L61
\reference{}Hurwitz, M., Bowyer, S., Bristol, W., Dixon, W.V., Dupuis, J., 
et al.  1998b, ApJ, 500, L1
\reference{}Jenkins, E.B.  1978a, ApJ, 219, 845
\reference{}Jenkins, E.B.  1978b, ApJ, 220, 107
\reference{}Jenkins, E.B.  1987, in Interstellar Processes, eds. D.J. 
Hollenbach \& H.A. Thronson, Jr.  (Dordrecht: Reidel), 533
\reference{}Jenkins, E.B.  1996, ApJ, 471, 292
\reference{}Lu, L., Savage, B.D., Tripp, T.M., \& Meyer, D.M.  1995, ApJ, 
447, 597
\reference{}Mihalas,D. \& Binney, J., 1981, Galactic Astronomy,
2$^{nd}$ edition (San~Francisco: Freeman)
\reference{}Morton, D.C.  1991, ApJS, 77, 119
\reference{}Reach, W.T., Koo, B-C., \& Helies, C.  1994, ApJ, 429, 672
\reference{}Reynolds, R.J.  1993, in Back to the Galaxy, ed. S. Holt \& 
F. Verter (New York: American Institute of Physics), 156
\reference{}Richter, P., Widmann, H., de~Boer, K.S., Appenzeller, I., 
G\"{o}lz, M., et al. 1998, A\&A, L9
\reference{}Robinson R.D., et al.  1998, PASP, 110, 68
\reference{}Savage, B.D., Bohlin, R.C., Drake, F.J., \& Budich, W.  1977, ApJ, 
216, 291
\reference{}Savage, B.D., Massa, D, \& Sembach, K.R. 1990, ApJ, 355, 14
\reference{}Savage, B.D. \& Sembach, K.R.  1991, ApJ, 379, 245 
\reference{}Savage, B.D. \& Sembach, K.R.  1996, ARA\&A, 34, 279
\reference{}Savage, B.D., Sembach, K.R., \& Cardelli, J.C.  1994, ApJ, 420, 183
\reference{}Savage, B.D., Sembach, K.R., \& Lu, L.  1997, AJ, 113, 2158
\reference{}Savage, B.D., Tripp, T.M., \& Lu, L.  1998, AJ, 115, 436
\reference{}Schectman, R.M., Povolny, H.S., \& Curtis, L.J.  1998, ApJ, 
504, 921
\reference{}Sembach, K.R. 1999, in the Proceedings of the Stromlo 
Workshop on High Velocity Clouds, eds. B.K. Gibson \& M.E. Putnam, 
ASP Conf. Series (San~Francisco: ASP), 243
\reference{}Sembach, K.R. \& Savage, B.D.  1992, ApJS, 83, 147
\reference{}Sembach, K.R., Savage, B.D, \& Lu, L. 1995, ApJ, 439, 672
\reference{}Sembach, K.R., Savage, B.D., \& Massa, D. 1991, ApJ, 372, 81
\reference{}Sembach, K.R., Savage, B.D., \& Tripp, T.M.  1997, ApJ, 480, 216
\reference{}Shelton, R.L. \& Cox, D.P.  1994, ApJ, 434, 599 
\reference{}Shull, J.M. \& Slavin, J.D.  1994, ApJ, 427, 784
\reference{}Slavin, J.D., Shull, J.M., \& Begelman, M.C.  1993, ApJ, 407, 83
\reference{}Snowden, S.L., Freyberg, M.J., Plucinsky, P.P., Schmitt, J.H.M.M., 
Truemper, J., et al. 1995, ApJ, 454, 643
\reference{}Soderblom, D.R., Hulbert, S.J., Leitherer, C., \& Sherbert, L.E.  
1994, Goddard High Resolution Spectrograph Instrument Handbook, v5 
(Baltimore: Space Telescope Science Institute)
\reference{}Sofia, U.J. \& Jenkins, E.B.  1998, ApJ, 499, 951
\reference{}Songaila, A. \& Cowie, L.L.  1996, AJ, 112, 335
\reference{}Spitzer, L.  1990, ARA\&A, 28, 71
\reference{}Spitzer, L.  1996, ApJ, 458, L29
\reference{}Steidel, C.C.  1990, ApJS, 72, 1
\reference{}Sutherland, R.S. \& Dopita, M.A.  1993, ApJS, 88, 253
\reference{}Tripp, T.M., Sembach, K.R., \& Savage, B.D.  1993, ApJ, 415, 652
\reference{}Wheelock, S.L., et al. 1994, IRAS Sky Survey Atlas: Explanatory 
Supplement (Pasadena: Jet Propulsion Laboratory)
\reference{}Widmann, H., de~Boer, K.S., Richter, P., Kr\"{a}mer, G., 
Appenzeller, I., et al. 1998, A\&A, 338, L1 

\newpage
\begin{center}
Figure Captions
\end{center}

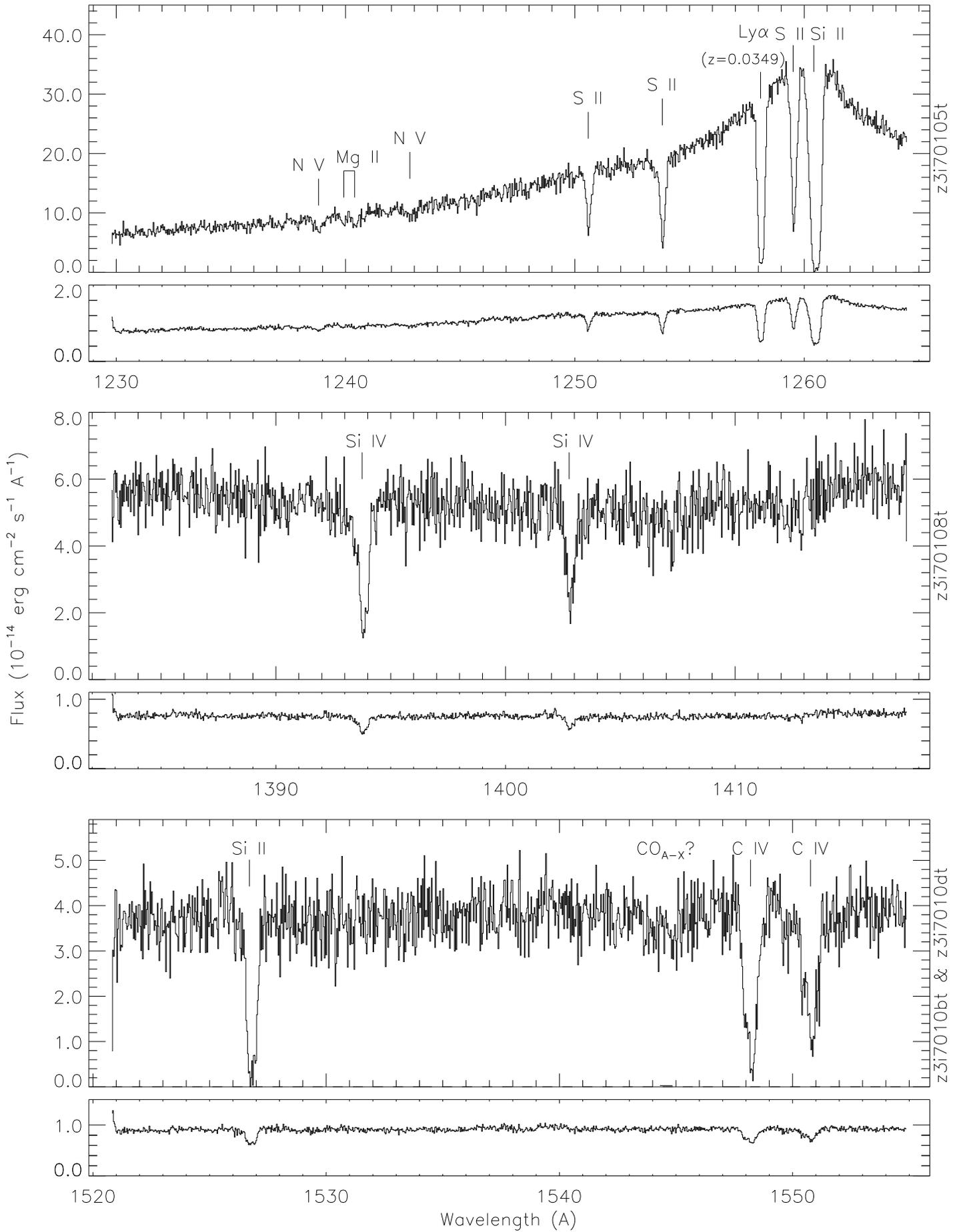
Fig. 1. --  GHRS G160M spectra of the \ion{N}{5}, \ion{Si}{4}, and \ion{C}{4}
spectral regions of ESO\,141-055. The data have S/N $\approx$ 13--21 per
resolution element and spectral resolutions (FWHM) of 14--18 \kms.
Error spectra (1$\sigma$) representing Poisson noise are shown below each data
spectrum. Interstellar lines discussed in the text are indicated.
The location of a redshifted Ly$\alpha$ line at 1258.126\,\AA\ 
($z$\,=\,0.03492) with W$_\lambda$\,$\approx$\,220 m\AA\ is also indicated.

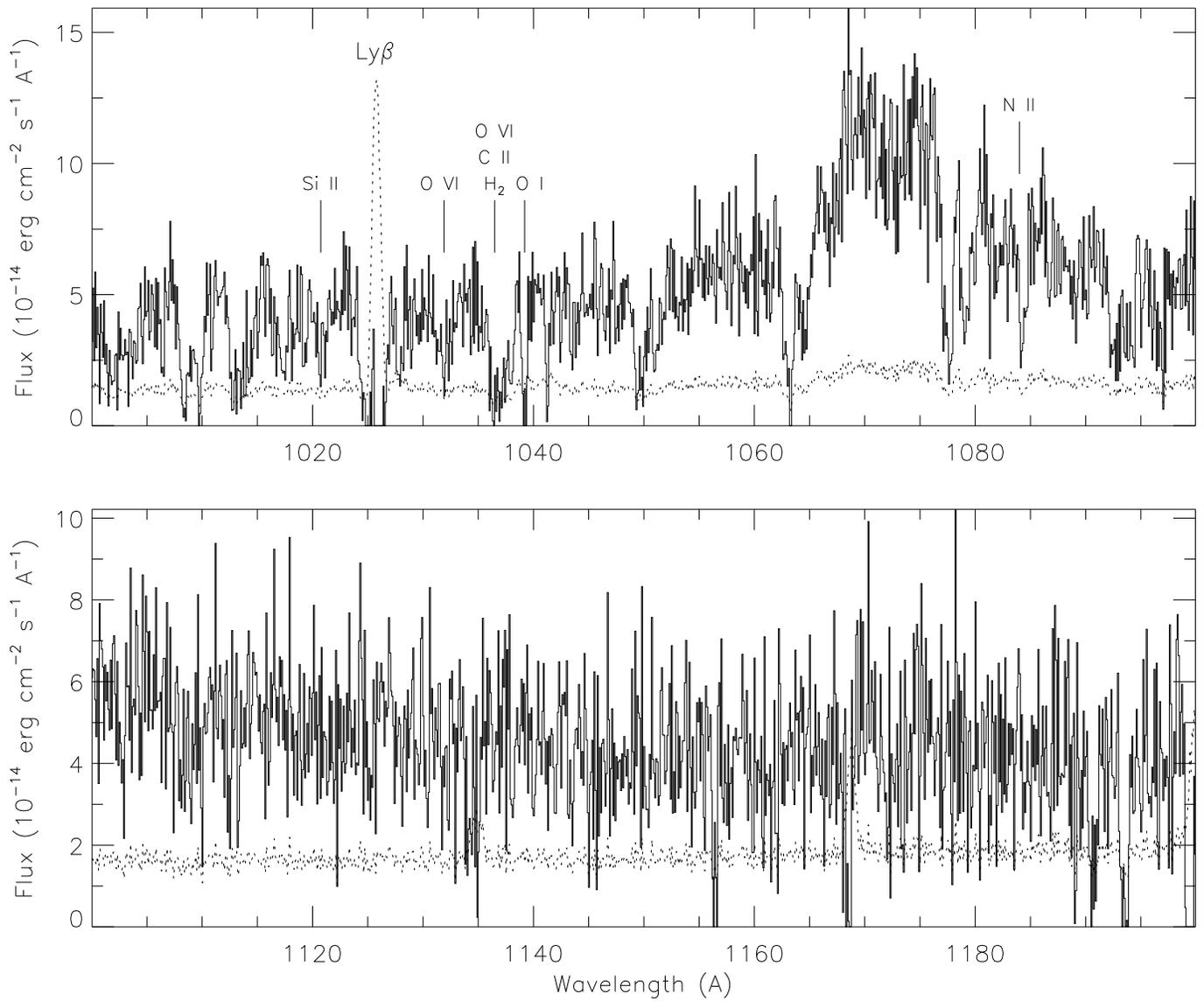
Fig. 2. -- A portion of the far-ultraviolet spectrum of ESO\,141-055 observed 
with the Berkeley spectrograph on the ORFEUS-II mission.  The data, which 
have been binned into 0.1\,\AA\ samples, have S/N $\approx$ 7 per 0.3\,\AA\
(100 \kms) resolution element.   Airglow features have been removed.  The 
dashed line is a 1$\sigma$ error vector.  Data below 1000\,\AA\ are not shown
since the S/N is very low ($<$3 per resolution element).  

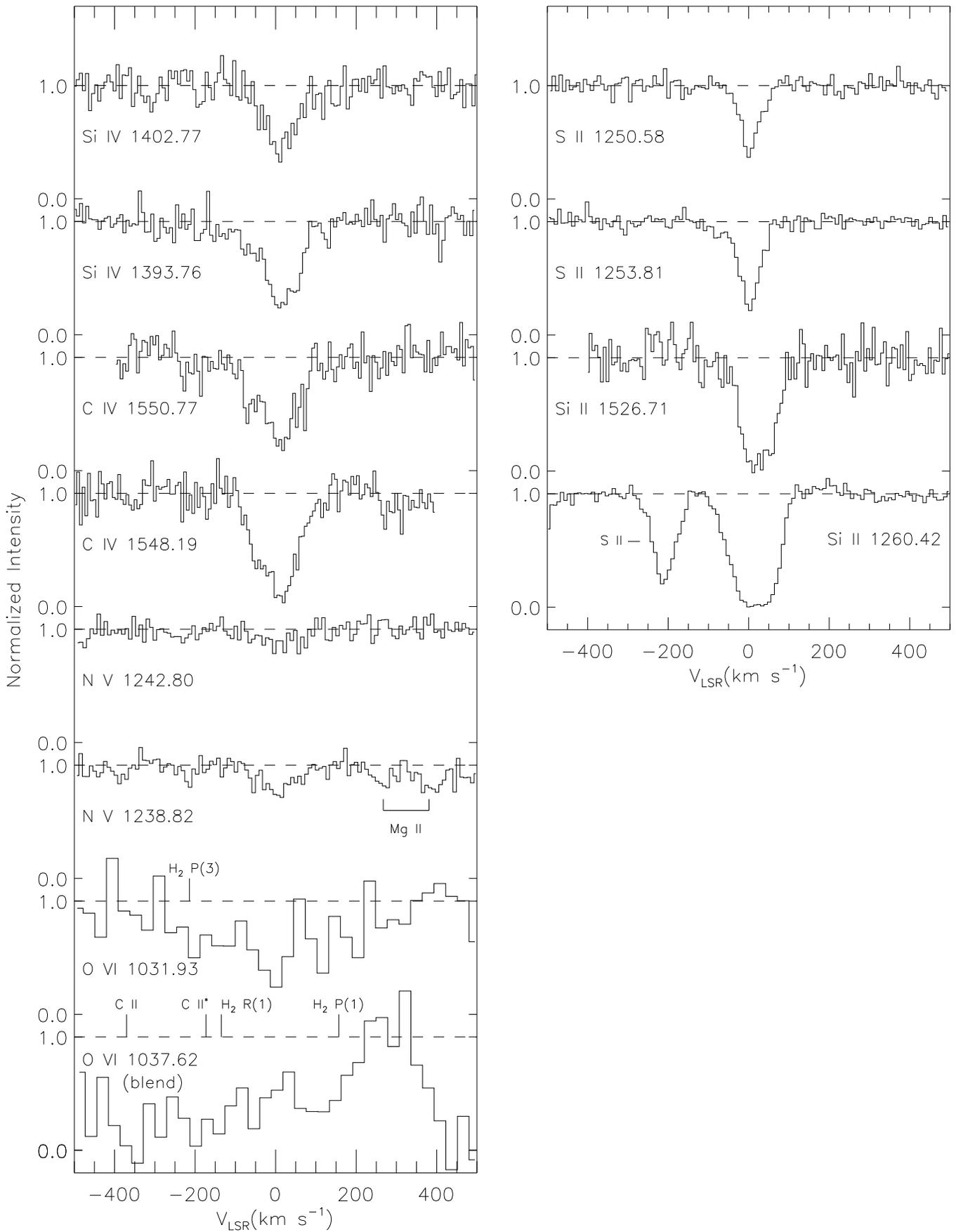
Fig. 3. --  Normalized absorption profiles of the high ionization lines 
toward ESO\,141-055 observed with the GHRS (\ion{Si}{4}, \ion{C}{4}, 
\ion{N}{5}) and ORFEUS-II (\ion{O}{6}).     
The \ion{Si}{2}\,$\lambda$1260.42 line includes blended 
\ion{Fe}{2}\,$\lambda$1260.53 absorption near +26 \kms. The 
\ion{O}{6}\,$\lambda$1037.62
absorption is blended with absorption by \ion{C}{2}\,$\lambda$1036.34,  
(--370 \kms), \ion{C}{2}$^*$\,$\lambda$1037.02 (--173 \kms), and the 5--0 R(1) 
and P(1) lines of H$_2$ near 1037.15\,\AA\ (--135 \kms) and 1038.16\,\AA\
(+157 \kms).  The \ion{O}{6}\,$\lambda$1031.93 absorption is blended
with the H$_2$ (6--0) P(3) line at 1031.19\,\AA\ (--214 \kms).  Several
\ion{Si}{2} and \ion{S}{2} lines observed with the GHRS are shown in the right
panel for comparison with the high ion lines.

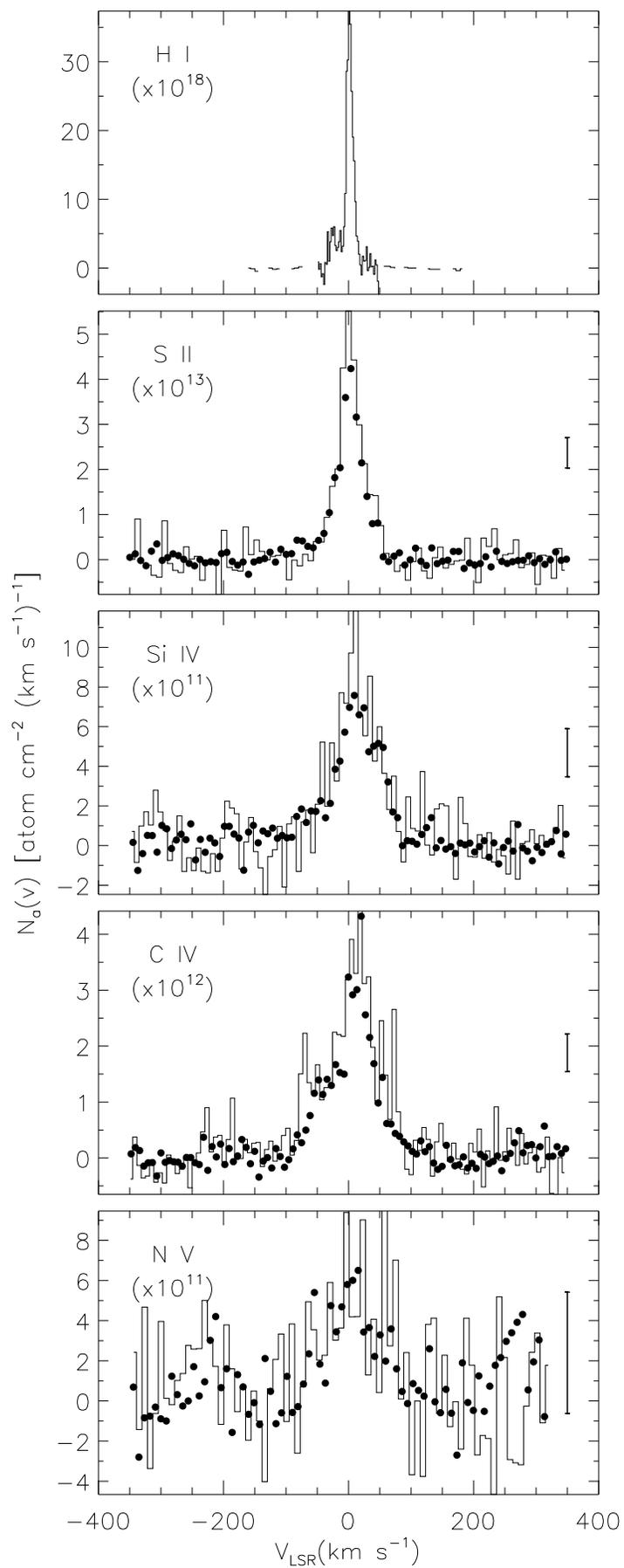
Fig. 4. -- {\it Top Panel}: \ion{H}{1} 21\,cm emission profile composite from 
Colomb et al. (1980) for $|$\vlsr$|$ $\le$ 50 \kms\ ($\Delta$v = 2 \kms, 
beamwidth (HPBW) = 30\arcmin) and from Cleary et al. (1979) for 
$|$\vlsr$|$ $\ge$ 50 \kms\ ($\Delta$v = 7 \kms, beamwidth (HPBW) = 48\arcmin).
{\it Lower Panels}: Apparent column density profiles for the \ion{S}{2}, 
\ion{Si}{4}, \ion{C}{4}, and \ion{N}{5} lines observed with the GHRS.  The
histograms are the profiles for the weaker member of each 
doublet/multiplet, and the filled points are the profiles for the stronger
member.  Systematic underestimates of $N_a(v)$ by the stronger lines at
some velocities (typically, the line cores) indicate that modest saturation 
corrections are necessary to recapture the total column densities of 
\ion{S}{2}, \ion{Si}{4}, and \ion{C}{4} (see text). Typical errors associated 
with the individual values of $N_a(v)$ are shown as vertical error bars at 
the right of each panel.

Fig. 5. -- Enlarged view of the 1000--1100\,\AA\ spectral region of the 
ESO\,141-055 ORFEUS-II data shown in Figure~2.  The data is shown as a 
histogrammed spectrum.  The solid smooth curve is the absorption spectrum
predicted by the neutral cloud model described in \S3.2.  Molecular 
hydrogen and atomic features arising in the \ion{H}{1} gas
are labeled. (The H$_2$ lines are indicated by their rotational and 
vibrational levels.)  The model cloud has log\,$N$(\ion{H}{1})\,=\,20.54,
log\,$N$(H$_2$)\,=\,19.00, and a solar mix of heavy elements.
Atomic features due to \ion{Ar}{1} and \ion{Fe}{2} that are not seen or are 
weaker than the model predictions are indicated under the spectrum with 
asterisks.  Absorption features due to ionized gas (e.g., 
\ion{O}{6}\,$\lambda$1031.93 and \ion{N}{2}\,$\lambda$1083.99) are present 
in the spectrum.  There is an unidentified feature at 1041.2\,\AA.

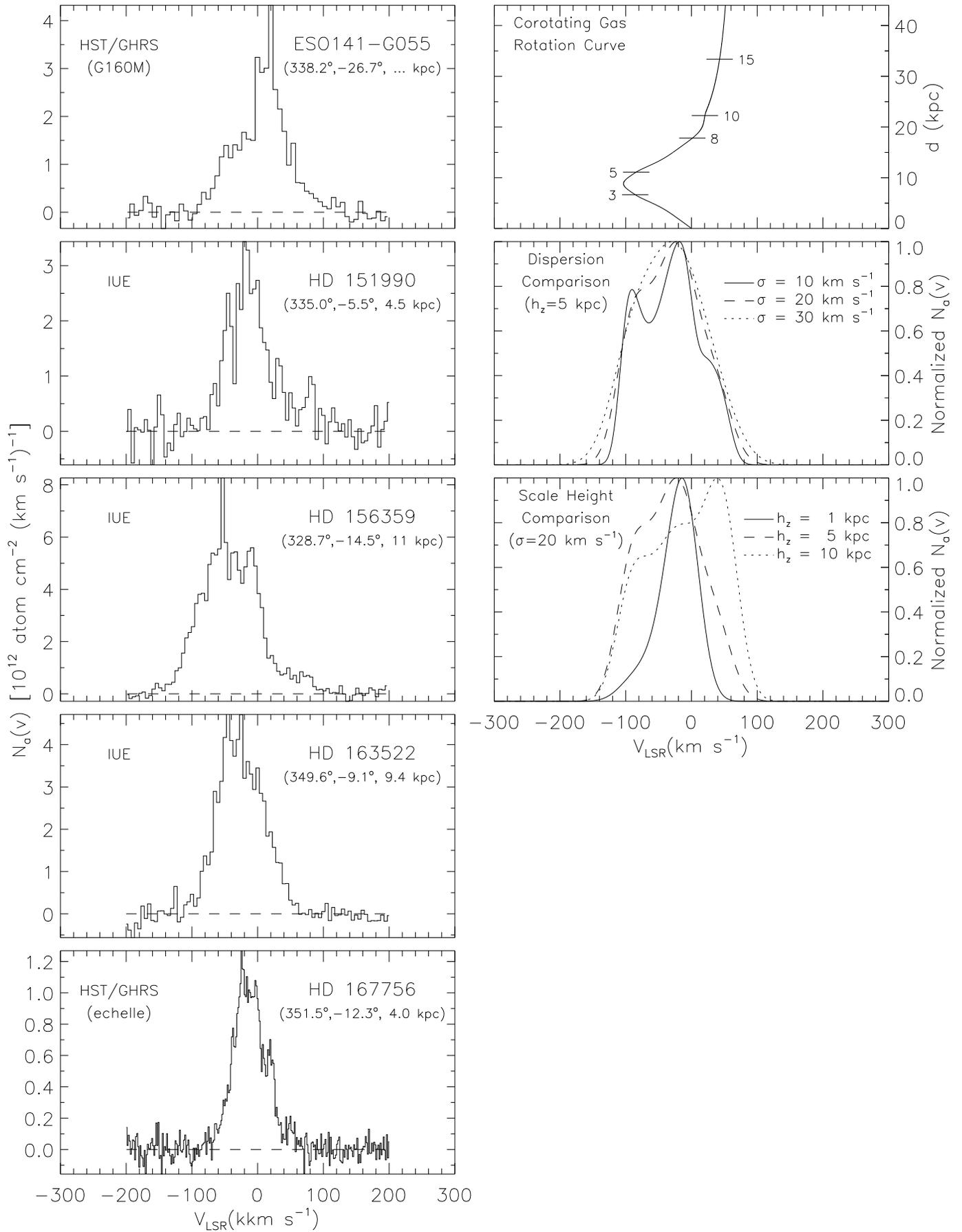
Fig. 6. -- {\it Left panels}: \ion{C}{4} apparent column density profiles
for the ESO\,141-055 sight line and the sight lines to four stars located in 
the inner regions of the Galaxy.  The Galactic coordinates and distances
of each object are indicated under the object name.  The $N_a(v)$ 
profiles for the stellar sight lines are from IUE or GHRS data presented in
previous work cited in \S4.2.  {\it Right panels}: These three panels 
contain the Galactic rotation curve and predicted apparent column density 
profiles for the
ESO\,141-055 sight line assuming a smooth exponential gas distribution away
from the Galactic plane.  The rotation curve is from Clemens (1985); the 
horizontal ticks on this curve indicate the $|$z$|$-distance of the sight line
(in kpc) at a given velocity.  The model column density profiles are 
normalized to 
their peak values and are appropriate for cases in which either the 
velocity dispersion ($\sigma$) or vertical scale height ($h_z$) of the gas is 
allowed to vary.  The three values of the parameter being varied ($\sigma$
or $h_z$) are shown as solid, dashed, and dotted curves as identified in the 
panel legends.  The models assume that gas at all distances from the Galactic 
plane obeys the rotation curve shown in the top panel.

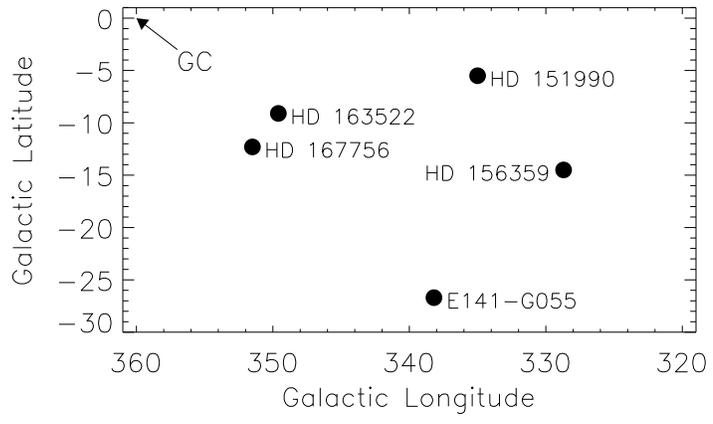
Fig. 7. -- Galactic locations of the sight lines for which $N_a(v)$ profiles 
are shown in Figure~6.  ``GC'' indicates the location of the Galactic center.

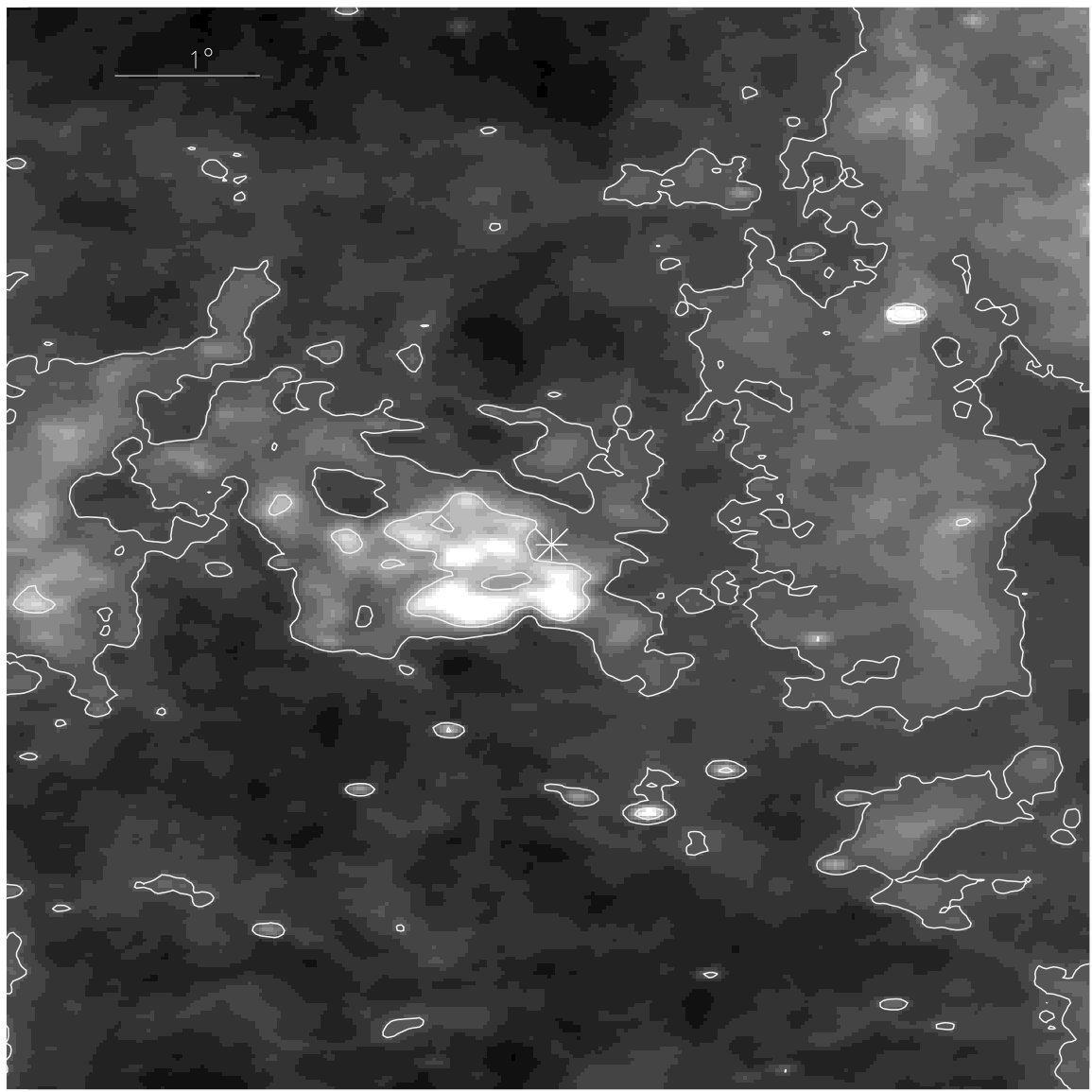
Fig. 8. -- A 7.5\degr\ $\times$ 7.5\degr\ IRAS 100$\mu$m map of the region of 
sky surrounding the ESO\,141-055 sight line (rectangluar equatorial coordinate
projection).  North is toward the top and east is to the left.  The location 
of the sight line is 
indicated by an asterisk.  These IRAS Sky Survey Atlas data have a spatial 
resolution of $\sim$2\arcmin.  The solid lines indicate 100$\mu$m
contour levels of 4, 8, and 12 MJy sr$^{-1}$. 

\end{document}


\newcommand{\degr}{$^{\circ}$}
\newcommand{\kms}{\,km\,s$^{-1}$}
\newcommand{\updag}{$^\dagger$}     
 
\begin{deluxetable}{lcccccccc}
\tablecaption{Equivalent Widths and Column Densities\tablenotemark{a}}
\tablehead{Ion & $\lambda$ & $f$ & $v_{min,max}$ & $<v_{LSR}>$ & FWHM & W$_\lambda$ & Log~$N_a$ & Log~$N$ \nl
 	       & (m\AA)    &   & (\kms)        & (\kms)    & (\kms) & (m\AA)    & (cm$^{-2}$) & (cm$^{-2}$)}
\startdata
\multicolumn{9}{c}{High Ions} \nl
\tableline
\ion{Si}{4} & 1402.770 & 0.2553 & $-$130, +140 & +12 & 92 & 262$\pm$24 & 13.90$\pm$0.04 & 13.96$\pm$0.06 \nl
      & 1393.755 & 0.5140 & $-$130, +140 & +13 & 110 & 409$\pm$22 & 13.84$\pm$0.04	 & (13.89$\pm_{0.09}^{0.12}$) \nl
\nl
\ion{C}{4}  & 1550.770 & 0.0952 & $-$130, +140 & +4 & 127 & 481$\pm$28 & 14.56$\pm$0.04 & 14.74$\pm$0.10 \nl
      & 1548.195 & 0.1908 & $-$130, +140 & +5 & 122 & 560$\pm$25 & 14.42$\pm$0.03	& (14.90$\pm_{0.20}^{0.71}$) \nl
\nl
\ion{N}{5}   & 1242.804 & 0.0782 & $-$130, +140 & +5 & 140 & 72$\pm$20 & 13.87$\pm_{0.14}^{0.10}$ & 13.82$\pm$0.10 \nl
      & 1238.821 & 0.1570 & $-$130, +140 & +4 & 137 & 112$\pm$17 & 13.77$\pm_{0.08}^{0.07}$ & (13.96$\pm_{0.23}^{0.61}$) \nl
\nl
\ion{O}{6}  & 1031.926 & 0.1329 & $-$150, +100 & $-$20: & 202: & 537$\pm$100 & $>$14.7 & 15.0$\pm$0.3\tablenotemark{b} \nl
      & 1037.617 & 0.0661 & ...                  & ... & ... & blend & ...	\nl
\tableline
\multicolumn{9}{c}{Low Ions} \nl
\tableline
\ion{Si}{2} & 1526.707 & 0.1270\tablenotemark{d} & $-$130, +130 & +20 & 100 & 531$\pm$25 & $>$14.30 & $>$14.30 \nl
      & 1260.422 & 1.1150\tablenotemark{e} & $-$130, +130 & +15  & 130 & $>$568\tablenotemark{c} & \nodata & \nodata \nl
\nl
\ion{S}{1} & 1401.514 & 0.0149\tablenotemark{f}  & $-$100, +90 & \nodata & \nodata & $<$32 & $<$14.09 & $<$14.09 \nl
\nl
\ion{S}{2} & 1250.584 & 0.0054 & $-$100, +90 & +3 & 45	 & 135$\pm$13 & 15.38$\pm$0.05 & 15.45$\pm$0.06 \nl
      & 1253.811 & 0.0109 & $-$100, +90 & +3 & 45	 & 212$\pm$10 & 15.31$\pm$0.03 & (15.39$\pm_{0.09}^{0.11}$) \nl
      & 1259.519 & 0.0162 & $-$100, +90 & ... & ... & $>$221\tablenotemark{c} & $>$15.20\tablenotemark{c} & \nodata
\enddata
\tablenotetext{a}{The entries in this table include the ion observed, 
vacuum wavelength from Morton (1991), $f$-value from Morton (1991) 
unless otherwise specified, approximate velocity range, average velocity, 
full width at half maximum intensity, equivalent width, integrated apparent 
column density, and adopted column density based on the values of $N_a$ 
as described in the text.  The numbers in parentheses in the final column are 
single-component Doppler-broadened curve-of-growth estimates based on the 
listed equivalent widths.  The associated b-values are b(\ion{Si}{4}) = 48$\pm_{12}^{30}$ \kms,
b(\ion{C}{4}) = 32$\pm_{8}^{6}$ \kms, b(\ion{N}{5}) = 14$\pm_{7}^{...}$ \kms, and b(\ion{S}{2})
= 27$\pm_{7}^{14}$ \kms.  All errors are 1$\sigma$ estimates.  All limits are 2$\sigma$ estimates.}
\tablenotetext{b}{This \ion{O}{6} column density was derived from the listed equivalent width,
assuming a single component Doppler-broadened curve-of-growth with a b-value of 57$\pm$8 \kms, as determined by
fitting a single Gaussian component to the \ion{N}{5}\,$\lambda$1238.82 line.}
\tablenotetext{c}{Interstellar \ion{Si}{2}\,$\lambda$1260.42 and \ion{S}{2}\,$\lambda$1259.52 occur near
the peak of the Ly$\alpha$ emission of ESO\,141-055.  
The ESO\,141-055 continuum curvature is uncertain at these
wavelengths, and therefore the 
listed values of W$_\lambda$ and $N_a$ for these lines should be considered lower limits.}
\tablenotetext{d}{$f$-value from Schectman et al. (1998).}
\tablenotetext{e}{$f$-value from Dufton et al. (1983).}
\tablenotetext{f}{$f$-value from Federman \& Cardelli (1985).}

\end{deluxetable}

\clearpage
\begin{deluxetable}{lcccccc}
\tablecaption{Projected Column Densities (Log\,$N \sin |b|$) for the High Ions}
\tablehead{& ESO\,141-055 & HD\,36402 (LMC)\tablenotemark{a} & PKS\,2155-304\tablenotemark{b} 
& 3C\,273\tablenotemark{c}  & Galactic Average\tablenotemark{d} \nl
& ($b$ = --26.7\degr) & ($b$ = --33.0\degr) & ($b$ = --52.2\degr) & ($b$ = +64.4\degr) }
\startdata
\ion{Si}{4} & 13.61$\pm$0.06 & 13.30$\pm$0.14 & \nodata & 13.74$\pm$0.04 & 13.56$\pm^{0.07}_{0.08}$\nl
\nl
\ion{C}{4}  & 14.39$\pm$0.10 & 13.90$\pm$0.03 & 14.01$\pm$0.04 & 14.45$\pm$0.03 & 14.08$\pm^{0.07}_{0.08}$\nl
\nl
\ion{N}{5}  & 13.47$\pm$0.10 & \nodata & 13.62$\pm$0.13 & 13.83$\pm$0.06 & 13.40$\pm^{0.09}_{0.12}$\nl
\nl
\ion{O}{6}  & 14.65$\pm$0.30 & 14.51$\pm^{0.15}_{0.24}$ & $<$14.23 & 14.80$\pm^{0.11}_{0.15}$ & ({\it 14.59})\nl

\enddata
\tablenotetext{a}{Values for HD\,36402 are from Sembach \& Savage (1992) and Widmann et al. (1998).}
\tablenotetext{b}{Values for PKS\,2155-304 are from Savage et al. (1997) and Hurwitz et al. (1995).}
\tablenotetext{c}{Values for 3C\,273 are from Sembach et al. (1997) and  Hurwitz et al. (1998b).}
\tablenotetext{d}{These Galactic averages for \ion{Si}{4}, \ion{C}{4}, and \ion{N}{5} (Savage et al. 1997)
are based upon 8--12 AGN/QSO sight lines and $\sim$30 sight lines toward OB stars in the 
Galactic disk and low halo. The \ion{O}{6} value is an average for the four sight lines listed, with the upper
limit for PKS\,2155-304 treated as a detection.}
\end{deluxetable}

\clearpage
\begin{deluxetable}{lccccccc}
\tablecaption{Comparison of Observed ionic Ratios to Theory}
\tablehead{Ratio & \multicolumn{3}{c}{Observed\tablenotemark{a}} & $<$Disk-Halo$>$\tablenotemark{b} & Mixing & Cooling & Conductive \nl
                 & ESO\,141-055 & PKS\,2155-304 & 3C\,273 & & Layers\tablenotemark{c} & Fountain\tablenotemark{d} & Interface\tablenotemark{e}}
\startdata
N(\ion{C}{4})/N(\ion{Si}{4}) & 6.0$\pm$1.5   & \nodata     & 4.8$\pm$0.5    & 5.1$\pm$0.6  & 1.6--4.5 & 3.0--4.7 & 20--26   \nl
\nl
N(\ion{C}{4})/N(\ion{N}{5})  & 8.3$\pm$2.4   & 2.5$\pm$0.7 & 4.0$\pm$0.6    & 4.2$\pm$0.6  & 9.1--28  & 2.2--6.8 & 1.8--2.6 \nl
\nl
N(\ion{C}{4})/N(\ion{O}{6})  & 0.55$\pm$0.30 & $<$0.6      & 0.44$\pm$0.13  & $_{0.3-1.7~Halo}^{0.1-0.3~Disk}$    & 1.2--3.4 & 0.1--0.5 & 0.2--0.4
\enddata
\tablenotetext{a}{Observed ionic ratios for ESO\,141-055 (this paper), PKS\,2155-304 
(Hurwitz et al. 1995; Savage et al. 1997),
and 3C\,273 (Sembach et al. 1997; Hurwitz et al. 1998b).}
\tablenotetext{b}{Average disk and halo sight line ratios from Sembach et al. (1997)
for N(\ion{C}{4})/N(\ion{Si}{4}) and N(\ion{C}{4})/N(\ion{N}{5}).  The ranges for 
N(\ion{C}{4})/N(\ion{O}{6}) are from Spitzer (1996).}
\tablenotetext{c}{Turbulent mixing layer ratios from Slavin et al. (1993) for a 
post-mixed gas temperature of 3$\times$10$^5$ K, solar abundances,  and an entrainment 
velocity of 25--100 \kms.} 
\tablenotetext{d}{Radiative cooling ratios for a gas heated to $\sim$10$^6$\,K and 
allowed to cool with self-photoionization within the flow included (Benjamin
\& Shapiro 1999).}
\tablenotetext{e}{Magnetized conductive interface ratios from Borkowski et al. 
(1990) for a front with an age of 2.5$\times$10$^5$ years and magnetic field inclinations 
between 60\degr--0\degr\ relative to the front normal direction.}
\end{deluxetable}